\documentclass[preprint]{aastex}
\begin{document}

\newcommand\gusten{G\"{u}sten}
\newcommand\kms{km s${}^{-1}$}
\newcommand\ab{$\sim$}
\newcommand\yz{Yusef-Zadeh}
\newcommand\p{$\pm$}
\newcommand\ad{$\alpha$,$\delta$$_{(B1950)}$}
\newcommand\beam{beam$^{-1}$}
\newcommand\x{$\times$}
\newcommand\T{T$_e^*$}
\newcommand\pc{pc$^{-1}$}

\title{The Molecular Component of the Galactic Center Arched Filament H II
Complex: OVRO Observations of the CS(J=2$-$1) Line}
\author{Cornelia C. Lang\altaffilmark{1}, W. M. Goss\altaffilmark{2}, Mark Morris\altaffilmark{3}}
\altaffiltext{1}{Dept. of Physics \& Astronomy, Van Allen Hall, University of Iowa, Iowa City, IA, 52242; email: clang@astro.umass.edu}
\altaffiltext{2}{National Radio Astronomy Observatory, Box 0, Socorro, NM 87801}
\altaffiltext{3}{Department of Physics \& Astronomy, 8371 Math Sciences
Building, University of California, Los Angeles, CA 90095-1562}

\begin{abstract}
The Owens Valley Radio Observatory (OVRO) millimeter array was used to make observations of the CS(2-1) line (at 97.981 GHz) arising from the G0.07+0.04 region of the ``$-$30 \kms'' molecular cloud near the Galactic center with a spatial resolution of \ab8\arcsec. The ionized edges of this cloud forms the Arched Filament HII regions which are ionized by the adjacent hes stellar cluster. The OVRO data were combined with single-dish data obtained at the 30-m IRAM telescope by Serabyn \& \gusten~(1987). A comparison of this CS(2-1) data and the H92$\alpha$~recombination line data of Lang, Goss \& Morris (2001) reveals that the ionized and molecular gas are physically related, but that their velocities in this region differ by up to 35 \kms. This difference in velocity can be understood if the gas that gave rise to the G0.07+0.04 HII region has been fully ionized. An overall comparison of the molecular and ionized gas across the entire $-$30 \kms~cloud based on the single dish CS(2-1) data and the H92$\alpha$~line data illustrates that such differences in velocity between the ionized and molecular gas are common and that the geometrical arrangement of these components is complicated. Much of the ionized gas resides on the near side (to the observer) of the molecular cloud; however, in several regions, some molecular material must lie in front of the HII region. The Arches stellar cluster therefore appears to be located in the midst of the molecular clouds such that some of the near-side cloud surfaces along our line of sight have not been exposed to the ionizing radiation.

\end{abstract}
\clearpage

\section{Introduction}
The interplay between components of the remarkable Galactic Center Radio Arc remains one of the outstanding issues in understanding the interstellar medium in this region. The Radio Arc, which lies \ab30 pc in projection from the center of the Galaxy, SgrA$^*$, was first revealed in detail with the Very Large Array (VLA) of the National Radio Astronomy Observatory\footnotemark\footnotetext{The National Radio Astronomy Observatory is a facility of the National Science Foundation, operated under a cooperative
agreement with Associated Universities, Inc.} over seventeen years ago (Yusef-Zadeh et al. 1984). The Arc consists of both thermal and nonthermal structures apparently interacting with each other, but the nature of the physical connections is not well understood. 
The prominent non-thermal filaments (NTFs) oriented perpendicular to the Galactic plane define the striking linear morphology of the Radio Arc. Eight systems of similar NTFs have been discovered within 250 pc of the Galactic center: typically, they extend for up to 60 pc in length, but are very narrow structures ($<$0.1 pc). The NTFs also show strong linear polarization, have intrinsic magnetic field orientations aligned along their long axis, and have magnetic field strengths estimated at 0.1-1 mG (Yusef-Zadeh \& Morris 1987a; Tsuboi et al. 1986; Yusef-Zadeh,
Wardle, \& Parastaran 1997; Lang et al. 1999a,b). The NTFs are thought
to trace a large-scale polodial magnetic field configuration in the inner Galaxy (Morris 1994). However, the origin of the relativistic particles in
these synchrotron NTFs and the mechanism for particle acceleration remain unclear. 

In addition, two concentrations of ionized and molecular gas appear to be associated with the Radio Arc NTFs: (1) the Sickle (G0.18$-$0.04), located at the center of the Radio Arc and (2) the Arched Filaments, which define its western edge (Pauls et al. 1976, 1980; Yusef-Zadeh 1986; Serabyn \& \gusten~1987 (hereafter, SG87), 1991). VLA observations reveal these sources to be peculiar HII regions with extremely filamentary morphology as well as complex velocity structure, although the coherence scale for the thermal filaments is much smaller than that of the NTFs (Yusef-Zadeh \& Morris 1987b; Yusef-Zadeh, Morris, \& van Gorkom 1987; Lang, Goss \& Wood 1997; Lang, Goss \& Morris 2001 (hereafter LGM01)). Conventional photoionization was initially dismissed because of the presumed unlikely arrangement of stars required along these HII structures, and models for heating based on MHD and shock mechanisms were considered (Morris \& Yusef-Zadeh 1989; Serabyn \& \gusten~1991).  

Recently, however, high resolution infrared observations of this region have significantly advanced our understanding of the thermal radio structures in the Radio Arc. Two extraordinary clusters of young stars (the Quintuplet and Arches clusters) have been discovered, richly populated with O-stars and large numbers of Wolf Rayet and other highly evolved stellar types (Nagata et al. 1995; Cotera et al. 1996; Figer et al. 1995; Serabyn et al. 1998; Figer et al. 1999a,b). Both radio recombination line and far-infrared observations have shown that the Quintuplet, located at the center of curvature of the Sickle HII region, can provide adequate ionization of the ionized gas in the Sickle (Lang et al. 1997; Simpson et al. 1997). However, ionization of the Arched Filaments has been more difficult to understand in terms of a single cluster, in part due to the large extent and unusual morphology: the Arched Filaments cover 22 $\times$ 16 pc with an areal filling factor of at most 10\%. The uniformity of ionization properties over such a large region, as derived from Kuiper Airborne Observatory (KAO) far-infrared observations, made the Arched Filaments even more puzzling, as it would be very difficult to distribute stars near the cloud surface in such a way as to account for that uniformity (Colgan et al. 1996). 

The recent recombination line study of LGM01, coupled with the recognition that the Arches stellar cluster generates a substantial ionizing flux ($\sim$2 x 10$^{52}$ s$^{-1}$), has revealed that this cluster adequately accounts for the ionization of the thermal Arched Filaments. The uniformity of physical conditions in the ionized gas is likely to be due to the large distance at which the cluster is located from the Arched Filaments (as much as 20 pc along the line of sight). Several outstanding questions remain in understanding this complex, most notably, the physical arrangement of the ionized, molecular and stellar components. Detailed comparisons of the distribution and kinematics of the ionized and molecular gas may offer some insight into the interactions of these components.

Furthermore, the nature of the interaction between the thermal structures and the magnetic filaments is unclear. Serabyn \& \gusten~(1991) first noted that the molecular material near the Sickle HII region may be physically associated with the linear NTFs in the Radio Arc, and that the electrons in the ionized gas may be accelerated to relativistic energies via magnetic reconnection. Follow-up interferometric observations of the CS(2-1) line emission in this cloud showed that the molecular gas is distributed in discrete clumps corresponding well to positions where both NTFs and ionized gas are present, and where the NTFs undergo striking changes in brightness and continuity (SM94). These authors propose that magnetic field reconnection occurs between the cloud field and the field in the magnetic NTFs, thereby causing the acceleration of some of the particles to relativistic energies. The relativistic particles then are constrained to move along the NTFs as they emit synchrotron radiation. This model relies on the presence of three elements: (1) a molecular cloud with a surface moving at a relatively large velocity compared to the intercloud medium; (2) a turbulent, ionized surface on this cloud to provide electrons and sufficient mixing between the cloud and intercloud medium; and (3) a magnetic field in the partially ionized molecular cloud which has a different orientation than the ambient magnetic field. In fact, far-infrared polarization observations of the gas underlying the Sickle have confirmed that these clouds possess internal magnetic fields aligned along the Galactic plane and therefore perpendicular to the NTFs (Morris \& Serabyn 1996). In the region of the Arched Filaments, however, the relationship between the molecular material and the Northern Thread NTF has not been explored, therefore providing an additional site to test the model of SM94.

In order to investigate (1) the arrangement of the ionized, molecular and stellar components in the Arched Filament complex, and (2) the nature of the intersection of the ionized and molecular gas with the Northern Thread NTF, 
 we carried out high-resolution millimeter observations with the OVRO millimeter array of the molecular gas in a portion of the Arched Filaments known as G0.07+0.04. This portion of the Arched Filament H II complex is intersected by the Northern Thread NTF (Lang et al. 1999). The three
elements required by the model of SG94 (molecular gas, ionized gas, and an NTF) are thus present. Figure 1 shows a diagram of the Arched Filaments with the field of view of the OVRO observations and the prominent sources labelled. In addition, we present a careful comparison of the morphology and velocity structure of the ionized and molecular gas across the Arched Filament region using the H92$\alpha$ line observations of LGM01 and previously published single-dish molecular line observations of SG87 to provide constraints on the geometrical configuration of the components. Details of the OVRO observations, the data reduction, and the combination of the OVRO and 30-m data of SG87 are summarized in $\S$2; the 3.4 mm continuum and the CS (2-1) line results are presented in $\S$3, and $\S$4 provides a discussion of the results. We assume throughout a distance to the Galactic center of 8.0 kpc (Reid 1993).

\section{Observations \& Data Reduction}
Observations of the W1 Arched Filament region in the 3.4 mm continuum
and CS(2-1) line were made with the six-element array at OVRO in 1998 October, November \& December. Three telescope configurations
(equatorial, low, and high) were used, with baselines ranging from 15 to 220 m.
Four fields with a primary beam of \ab60\arcsec~were observed in a mosaic
pattern, with a spacing of 30\arcsec. The resulting mosaic covers an area of
approximately 2\arcmin~\x~2\arcmin, centered on the position \ad=17 42 24.0,
$-$28 49 40. In this paper we use B1950 coordinates since the OVRO
data are combined with single dish data observed in the B1950 reference frame. 
The total integration time on each field was \ab4 hours. NRAO530, 3C273 and Neptune
were used for gain, passband and absolute flux calibration, respectively. 
The data were calibrated using the MMA package (Scoville et al. 1993), and the
mosaicking was carried out with the maximum entropy method of deconvolution
implemented in the MIRIAD routine {\it mosmem} (Cornwell \& Braun 1988; Sault,
Stavely-Smith \& Brouw 1996). The resolution of the final mosaicked image is 8\farcs1 $\times$ 4\farcs9, PA=$-$10\fdg8.  The CS(2-1) line data were observed at a rest
frequency of 97.981 GHz, with 64 channels of 0.5 MHz width, corresponding to a velocity resolution of 1.53 \kms, and a velocity coverage of \ab96 \kms. The line was centered on $-$40 \kms.  Simultaneous 3.4
mm continuum observations were made with a bandwidth of 1 GHz.

\subsection{Addition of Single Dish Data}

The largest spatial scale to which the OVRO interferometer is
sensitive is 20\arcsec, corresponding to the shortest baseline of 15 m
at 3.4 mm. Therefore, more extended structures are not detected. In order to recover
the missing flux density, the total-power measurements from single dish
observations of this region have been added. Single-dish observations of the
CS(2-1) line in the $-$30 \kms~molecular
cloud were carried out with the IRAM 30-m telescope by SG87. Spectra in the vicinity of the Arched Filament complex were
obtained at regular grid spacings of 18\arcsec, and imaged with a single-dish beamsize of \ab25\arcsec. These observations were centered
at V$_{LSR}$=0 \kms, using a 512-channel filterbank with 1-MHz resolution,
which corresponds to a velocity resolution of 3.06 \kms.
\clearpage
Since there is reasonable overlap between the shortest spacings of the OVRO
interferometer (4 k$\lambda$) and the diameter of the 30-m antenna (8
k$\lambda$), the linear technique of ``feathering'' single dish and
interferometer data is appropriate. This method requires that single dish data
be a good representation of the object at low spatial frequencies, and that the
interferometer mosaic is a good representation at mid-to-high spatial
frequencies. The feathering technique can be carried out using the MIRIAD task
{\it immerge}. We input deconvolved and restored single dish and
interferometeric images with the same velocity resolution and spatial grid. {\it
Immerge} first transforms the images into the Fourier plane, where the data are
combined. The single dish data are given unit weight, and the low spatial
frequencies of the interferometer data are downweighted in the Fourier plane
with a taper such that a combination of the single dish and interferometer data
results in an image with a gaussian beam equal in diameter to the beam of the input interferometric mosaic image.

In the case of the 30-m and OVRO millimeter array data, the flux densities in
the overlap region (4-8 k$\lambda$) agree at the 10\% level.  To prepare the
images to match exactly, the interferometer data
were smoothed to the velocity resolution of the single dish
observations (1 MHz, or 3.06 \kms), and deconvolved using the {\it
mosmem} algorithm as above, then convolved with a FWHM=8\arcsec~taper to give a 
beam of 8\farcs8 \x~7\farcs8, PA=$-$17\fdg8. A cube was
created from the single dish data with the same number of channels, assuming a
gaussian beam of 25\arcsec, and the units were converted to Jy \beam, using a
conversion factor of 4.7 Jy/K (Mauersberger et al. 1989). The resulting combined
image has a resolution the same as the deconvolved interferometer data.

\section{Results}
\subsection{3.4 mm Continuum}

Figure 2 shows the 3.4 mm continuum emission from the OVRO millimeter array
observations overlaid on the 3.6 cm continuum emission from the VLA observations
of LGM01. The resolution of the uniformly-weighted 3.4 mm continuum
image is 9\farcs2\x~4\farcs6, PA=$-$3\fdg2, with an rms noise of 6 mJy \beam.
There is close correspondence between the
continuum morphology at both frequencies. The 3.4 mm continuum traces
the brightest part of the W1 filament (G0.07+0.04), although the 3.4 mm peak 
(35 mJy \beam) is slightly offset from
the 3.6 cm continuum peak. The flux density at 3.4 mm is considerably lower than
at 3.6 cm, probably owing to the fact that much of the extended structure at 3.4
mm ($>$ 20\arcsec) is resolved out. A visual comparison of the distributions in
Figure 2 illustrates that the 3.4 mm observations are only sampling the most
compact peak of emission in G0.07+0.04. The 3.4 mm emission does not appear to
be dominated by dust emission in the H II region, which would produce a thermal
(rising) spectral index. Instead, the 3.6 cm to 3.4 mm spectrum is likely to be
flat, consistent with the 3.4 mm detection of the free-free emission
from the H II region with a minor contribution from dust emission.

\subsection{CS(2-1) Line Emission}
Figure 3 shows channel images of the CS(2-1) line from the OVRO observations alone
with a resolution of 8\farcs1 \x~4\farcs9, PA=$-$10\fdg8. Most of the emission
occurs in the velocity range $-$7 \kms~to $-$22 \kms, where the clump of 
emission appears to be slightly elongated in the E-W
direction. Figure 4a shows the CS(2-1) emission from the OVRO data
integrated over the velocity range of 0 to $-$50 \kms. As in the channel images
in Figure 3, the CS(2-1) clump in Figure 4a appears extended in the E-W
direction, with a forked structure on the W side.  In addition, there is a
compact emission structure present to the North of the main emission complex.

Figure 4b shows that the extended CS(2-1) emission has been recovered in the
combination of 30-m single dish and OVRO data. In particular, a diffuse feature
is apparent which apparently links the northern component to the southern emission clump.
Figure 5 shows spectra from (a) the OVRO only and (b) 30-m+OVRO images, integrated over a \ab1\arcmin~\x~1\arcmin~region which includes the CS(2-1) peak.
A comparison of these spectra illustrates that we only detect \ab50\% of
the total flux density with the OVRO interferometer compared to the resulting
combined image.  Figure 6 shows the spatial relationship between the ionized gas
in the Arched Filaments (shown in 8.3 GHz continuum contours from LGM01) and the CS(2-1) emission (shown in contours and greyscale). The peak of the CS(2-1) emission (at \ad=17 42
26.0, $-$28 50 00) is coincident with the 8.3 GHz continuum peak in G0.07+0.04.
At the locus of the Northern Thread, the CS(2-1) emission shows a minimum and is
bounded on either side by a ridge of emission elongated in a direction along the NTF. The velocity
structure of the CS(2-1) emission in this region is shown in Figures 7a and b.
Along four position angles (locations shown in Figure 7a), position-velocity
diagrams were constructed. In all cases, the velocities decrease to the
South. Along cuts B and D, the velocity has a nearly constant value (v\ab10 \kms), whereas along A and C there are velocity gradients up to 5 \kms~pc$^{-1}$. 

In order to understand the physical relationship between the molecular and
ionized gas, the velocities and lineshapes of these components can be compared.
CS(2-1) line profiles from these observations and the H92$\alpha$~emission from LGM01 were sampled at the same positions near G0.07+0.04 and are overlaid
in Figure 8. The most striking feature of the profile comparison is that at all
positions across the molecular gas clump there is an offset between the velocities of the H92$\alpha$~
and CS(2-1) lines of as much as 35 \kms. The H92$\alpha$~emission in
this region is characterized by velocities of \ab$-$50 \kms, whereas the CS(2-1)
emission has velocities between $-$10 and $-$15 \kms. Figure 8a represents the peak of the CS(2-1) emission, and there is good correspondence in position between the strong
H92$\alpha$~and CS(2-1) emission even though the velocities are quite discrepant.
In the western region of the clump (Figures 8c and d), the H92$\alpha$~lines are weak, but may be associated with components of the CS(2-1)
emission which appear at velocities of $-$40 \kms. In addition, the CS(2-1) spectra appear to have double peaks with velocity components of $-$20 to $-$25 \kms~and $-$10 to $-$15 \kms~in
Figures 8d, e, and f. These double CS(2-1) profiles are consistent with the two velocity components
detected in the molecular emission along cut B.

\section{Interpretation}
In this section, we discuss the physical relationships between the
components of the Arched Filaments complex: the $-$30 \kms~molecular cloud, the
ionized gas, the Arches stellar cluster, and the Northern Thread NTF. 
First we consider the implications of the OVRO+30-m
CS(2-1) observations of the southern region of the W1 Arched Filament, G0.07+0.04.  Second, we examine the relationship between the ionized and molecular gas over the
entire region of the Arched Filaments and attempt to elucidate the relative locations of the gaseous features and the Arches stellar cluster.

\subsection{Molecular Emission near G0.07+0.04}
\subsubsection{Relationship Between Molecular and Ionized Gas in G0.07+0.04}

The morphology and the velocities of the components in this particular region of the Arched Filaments complex show striking differences. Locally, the CS(2-1) emission defines a ridge oriented in the E-W direction (Figure 4), in contrast to the N-S elongated
emission in the ionized W1 filament. Differences in velocity between the ionized and molecular gas are in some cases up to 35 \kms, far exceeding the typical sound speed in an HII region (\ab10 \kms). These discrepancies raise the question of whether the ionized and molecular components in this region are physically related.  The velocity differences would suggest that these components are not related. Yet, the presence of some molecular material at velocities of $-$40 \kms~in this cloud complex (see SG87 and below in $\S$4.2.2) indicates that in the vicinity of G0.07+0.04, molecular material at $-$40 \kms~may have been almost fully ionized, leaving only the filamentary HII remnant G0.07+0.04. 
In other portions of the cloud complex, where complete ionization has not yet occured, the velocities and morphologies of the components appear to be more closely correlated than in the G0.07+0.04 region (see $\S$4.2). In addition, the peaks in the CS(2-1) and H92$\alpha$ lines occur at the same location (see Figure 6) despite the dramatically different morpohologies in this particular region. Overall, the highly unusual negative values of velocity characterizing both the ionized and molecular gas in this Galactic quadrant and the larger-scale similarities in morphology indicate that these components are likely to be physically related.

\subsubsection{Association Between the Molecular Gas and Northern Thread NTF}

A comparison of the 8.3-GHz radio continuum with the distribution of CS(2-1) 
emission (Figure 6) reveals that there is a concentration of molecular gas surrounding the
intersection of the NTF with the ionized W1 Arched Filament. The distribution of
molecular emission in Figure 6 resembles the clumps of CS(2-1) emission that SM94
observed at several positions in the molecular cloud associated with the NTFs of the Radio Arc and the
Sickle HII region. Therefore, we are detecting the close positional relationship between the NTF, 
the ionized gas, and the molecular gas that SM94 rely on for their particle accleration mechanism. 
In Figure 6, the Northern Thread NTF is essentially tangent to the clump displaying the strongest 
CS(2-1) peak, and the CS(2-1) emission is concentrated into two parallel ridges of extent 
\ab1\arcmin~(2.5 pc) on both sides of this NTF. However, this only represents one of the many positions in this cloud complex where the NTFs appear to intersect and it is difficult to demonstrate that these components are physically interacting. In addition, the existing high-resolution data only a small fraction of the total area of the $-$30 \kms~cloud. In order to determine whether 
the molecular clump$-$NTF relationship is present exclusively at positions where NTFs intersect 
the molecular gas in the Arched Filament Complex (as in SM94),
further high-resolution molecular observations of the entire cloud would be required.
 
\subsection{Relationship between Molecular and Ionized Gas throughout the Arched Filaments Complex}

Several previous studies have established that the ionized and molecular gas in the Arched Filament complex are closely associated (\gusten~\& Downes 1981; Yusef-Zadeh 1986; Bally et al. 1987; SG87). Of these papers, SG87 present a comparison of the molecular emission lines (CS(2-1)) lines and the ionized gas (from the H110$\alpha$~data of Yusef-Zadeh (1986)). The H110$\alpha$ data has a spatial resolution of 22\arcsec~and does not cover the full extent of the ionized structures, but confirms that the velocities of the gas components correspond well. Here, however, we provide a more complete examination of the morphology and kinematics of these gaseous components by comparing the data of SG87 with the recent H92$\alpha$~line data of LGM01 (spatial resolution of \ab13\arcsec) which covers the entire field of view of the Arched Filaments with adequate signal to noise.

\subsubsection{Morphology}

Images showing superpositions of CS(2-1) and H92$\alpha$ emission integrated over the same velocity ranges are presented in Figure 9. They illustrate clearly that the overall morphology of the $-$30 \kms~molecular cloud and the ionized filaments are very similar, with elongated structures running in an essentially N-S orientation along the Galactic plane. This elongation is consistent with the idea that these clouds have been stretched in this direction by the strong tidal gravitational fields present near the Galactic center (SG87; Morris et al. 1992). 

The spatial coincidence of the edges of the CS(2-1) and H92$\alpha$ emission regions is striking. SG87 first pointed out this correspondence (see Figure 5 of SG87), but here Figures 9a-c more clearly demonstrate this point. At several locations, the ionized gas lies exactly at the edge of the molecular gas. For example, in Figure 9c, the strongest peaks of the H92$\alpha$~line emission (shown in orange) along the E2 filament occur just at the edge of the molecular gas (shown in blue) at \ad=17 42 36.0, $-$28 49 15 and 17 42 36.0, $-$28 46 00. The morphology of the molecular gas also corresponds well to the curvature of the E2 ionized filament at these locations. In Figure 9b, the southern boundary of the
ionized portion of the W filaments seems to coincide exactly with the northern boundary of molecular gas in this velocity range. 

In both Figures 9 b and c, the prominent HII complex G0.10+0.12 has no direct molecular counterpart, but it is surrounded on either side spatially by molecular material; this anti-correlation suggests that this one portion of an otherwise continuous molecular cloud has been completely ionized. At other locations, the distribution of ionized and molecular gas correspond very closely, most notably in Figure 9c. SG87 suggest that the coincidence of the molecular gas along edges of ionized gas indicates that the cloud is being ionized from the side, whereas the close spatial correspondence of molecular and ionized features indicates that the ionization occurs on a cloud side facing the line-of-sight.

\subsubsection{Velocity Structure}

Beyond the correspondence in overall morphology, the ionized and molecular features also agree well in their central line velocities. The squares in Figure 10 represent positions at which the central velocity and lineshape of the CS(2-1) and H92$\alpha$ lines are compared. A sample of these profiles is shown in the panels labelled a-j. The majority of profiles shown in Figure 10 (a, b, c, d, e, f, h, \& j) show that the differences between velocities of at least one component of ionized and molecular gas are less than 10 \kms. For visual reference, the differences in velocities between the H92$\alpha$~and CS(2-1) components at each position are indicated as either blueshifted (blue squares) or redshifted (red squares) with respect to the H92$\alpha$~line central velocity. 

Double peaked CS(2-1) profiles are observed in several locations throughout this complex, and appear to be similar to the double-peaked structure in the H92$\alpha$ line profiles of LGM01. In the northern portion of the W filaments where W1
and W2 intersect (\ad=17 42 22.0, between $-$28 46 30 and $-$28 48 00), the
H92$\alpha$ profiles have double-peaked structure (see Figure 10c), which is attributed to the superposition of spatially-adjacent components of ionized gas which have different velocities. The same interpretation might be applied to the two components of molecular gas at $-$40 \kms~and $-$10 \kms~(Figure 10d), as this region is coincident with the intersection of these two filaments (W1 and W2).  

A further examination of the velocity structure of the ionized and molecular components was made by comparing the position-velocity structure of the ionized and molecular gas. Figure 11 shows a finding chart for the position-velocity diagrams shown in Figures 12a-d. The ABCD labels in Figure 11 correspond to Figures 12a (E1), 12b (E2), 12c (W1) and 12d (W2). In addition, three reference markers (1, 2, 3) are indicated in Figure 11 along each filament, corresponding to positions along each of the filaments in Figures 12a-d. 

Overall, these figures suggest that some anti-correlations exist on small scales in the velocities of the ionized and molecular emission, further supporting the idea that the morphology and arrangement of ionized and molecular gas are unusually complex. For example, at Position 2 in Figure 12a, the molecular gas is clearly double peaked (v\ab$-$17 and $-$37 \kms), whereas the spatially-coincident ionized gas has a velocity of v\ab$-$25 \kms, as if located between the two molecular features in velocity. Also, between Positions 1 and 2 in Figure 12b, the velocity of the ionized gas (v\ab$-$20 \kms) corresponds roughly to one of the peaks of molecular emission; yet there is a second component of molecular emission with a more negative velocity (\ab$-$40 \kms) and a clear lack of ionized gas in this velocity range. A comparison of the position-velocity diagrams for the ionized and molecular gas in G0.10-0.02 confirms that this strong HII complex likely represents the ionized portion of an otherwise continuous molecular cloud. The peak of molecular gas in the E2 filament occurs around position 1 (2\arcmin~along its length). Further along the E2 filament, near position 3 in Figure 12b, the molecular emission is much weaker, and the ionized gas has a peak around $-$20 \kms, suggesting that the molecular material in that velocity range has been mostly ionized; as Figures 9 b and c show, the molecular material may have a continuous distribution on either side of the ionized component. 

\subsection{Relative Placement of Components in the Arched Filament Complex}

The majority of H II regions are located at the edges of molecular clouds and therefore the velocity structure and morphology of the ionized gas are often are closely related to the underlying molecular
material (Israel 1978).  The distribution of ionized and molecular gas in the Arched Filaments complex may well be related to the ``champagne phase'' in the evolution of an H II region  (Tenorio-Tagle 1979).
This phase occurs when molecular material in physical contact with the H II region is evaporated by the ionization by nearby OB stars. In this paradigm, an H II region is assumed to have formed around a
cluster of stars located at the edge of a molecular cloud. The ionization front then breaks out of the molecular gas and begins to expand into the less dense,
surrounding interstellar medium, creating a stream of ionized gas which flows off of the molecular cloud surface. The result is that one side of the H II region becomes density-bounded, while the other side is ionization-bounded. 

The simple picture proposed by LGM01 for the arrangement of the cluster and gaseous features is based on the above idea: the Arches cluster lies near the edge of a dense molecular cloud, whose surface is thereby ionized, resulting in a flow of plasma away from the molecular cloud surface. For simplicity, the cluster is assumed to be on the near side (closest to the observer along the line of sight) of the cloud in LGM01. Therefore, the ``front'' side of the molecular cloud would be ionized. In fact, the Br$\gamma$ (2.166 $\mu$m) emission line study of Cotera et al. (2000) indicates that the ionized gas indeed lies on the near side of the molecular cloud. They observe Br$\gamma$ emission arising from one of the brightest regions of the Arched Filaments (G0.10+0.12) and conclude, based on a comparison of the theoretical and measured extinction, that the ionized gas in this region cannot be embedded within the molecular cloud and must lie on the side of the feature closest to the observer. 
\clearpage
In this picture, if the ionized gas is expanding from the near side of the cloud toward the observer, then the velocities of the H92$\alpha$ line emission might be consistently blueshifted from the molecular line emission. However, Figure 10 shows that the ionized and molecular components do not follow this simple prescription. Although there are regions of the Arched Filaments where the H92$\alpha$~lines appear to be blueshifted relative to the CS(2-1) emission (e.g. the profiles for portions of E1, E2 and much of W1: Figures 10a, b, \& g), there are also
regions where the H92$\alpha$~lines are redshifted relative to the CS(2-1) emission (i.e. profiles along W2: Figures 10d \& e). The inconsistency of the trend among the velocity profiles across the Arched Filaments suggests that the cloud may be comprised of quasi-independent gas features with different velocities and probably different line-of-sight distances, all of which are largely superposed in the plane of the sky to give the impression of a more unified structure. In addition, some portions of the Arched Filaments may be completely ionized and have no immediate molecular counterpart at a similar velocity (as in $\S$4.1.1); yet, at adjacent positions there is often evidence for molecular gas at similar velocities to the ionized gas, suggesting that the distribution of molecular gas is complex, and that ionization has not reached all surfaces equally. 

Previous lower resolution (\ab50\arcsec) VLA observations of HI absorption toward the Radio Arc can provide some geometrical constraints. The large HI absorption opacities toward the peaks of radio continuum emission in the westernmost Arched Filaments indicate that a layer of atomic gas lies in front of the ionized gas in W1 and W2 (Lasenby, Lasenby \& Yusef-Zadeh 1989). These authors also find that the HI absorption opacities are reduced toward the Eastern filaments, suggesting that at least {\it some} ionized material lies on the near side of E1 and E2, and that much of the molecular gas is located behind the ionized gas (Lasenby et al. 1989).

Thus, a more complex geometrical arrangement of the components than proposed by LGM01 is warranted. Such an arrangement must account for the unusual morphology of the ionized gas, the multiple-peaked structure of both the CS(2-1) and the H92$\alpha$ line profiles in various regions of the Arched Filaments, and the HI absorption results described above. If the stellar cluster resides on the near side of the molecular cloud, as originally proposed in LGM01, one would expect the entire front surface of the molecular cloud to be ionized. The presence of both HI and molecular gas on the near side of the cloud (as described above) are not consistent with this picture. Instead, the narrowness of the Arched Filaments and the presence of molecular gas and HI absorption on the near side of the cloud suggest that the molecular material has a finger-like distribution, of which only the protruding, narrow edges are ionized. The cluster is likely to be embedded within the distribution of molecular gas such that some of the cloud surfaces along our line of sight have not been exposed to the ionizing radiation.

Figure 13 shows a schematic of such an arrangement of the molecular, ionized, atomic and stellar components which satisifies some of the observed constraints. The dotted lines in Figure 13 represent positions in the Arched Filaments where multiple-peaked profiles can be explained by a superposition of features along the line of sight. These correspond to four of the profiles in Figure 10: sightline 1 (10j), 2 (10h), 3 (10g) and 4 (10d). In the Western filaments, it is apparent that some of the molecular and atomic gas has not been directly exposed to the ionizing radiation of the Arches cluster 
(most apparent on the western side of the W1 and W2 filaments, where the enhanced HI absorption opacity is present). In the Eastern filaments, most of the surfaces of the molecular cloud which are detected have been ionized; therefore the HI absorption opacity toward this region is more reduced than toward the Western side. A higher resolution (\ab15\arcsec) VLA study of the HI absorption in and around the Arched Filament complex is currently being carried out and will provide a more reliable picture of the line of sight location of these components.

\section{Conclusions}

Detections of the 3.4 mm continuum and the CS(2-1) line arising from a region
of the $-$30 \kms~molecular cloud at the Galactic center were made using the
Owens Valley millimeter array. These data were combined with the corresponding
total power data obtained at the 30-m IRAM telescope by SG87.
The following conclusions are made:

(1) Continuum emission at 3.4 mm was detected in the G0.07+0.04 region,
coincident with the peak of 3.6 cm continuum emission. The spectrum between 3.4
mm and 3.6 cm appears to be flat, consistent with a detection at 3.4 mm of the
free-free emission arising from the H II region and with a slight contribution from
dust.

(2) The combined OVRO+30-m integrated CS(2-1) emission image shows that overall the
molecular gas in the G0.07+0.04 region is distributed in a compact clump at the
intersection of the ionized filament and Northern Thread NTF. However, there is a pronounced decrease in the CS(2-1) emission along the exact trace of the Northern Thread. This might indicate that the NTF
and molecular gas are interacting, although the nature of the interaction is not clear.

(4) A comparison between the OVRO+30-m CS observations and the H92$\alpha$ data
of LGM01 in the region of G0.07+0.04 shows that the molecular and
ionized gas are physically related, although the velocities are separated in
some places by up to 35 \kms, indicating that one component of the molecular gas 
in this direction may have been fully ionized. 

(5) A larger-scale comparison between the 30-m CS observations and the
    H92$\alpha$ data from LGM01 also indicates that the
    molecular and ionized gas are closely associated across the entire
    region of the Arched Filaments, as was first discussed by SG87. 
    The velocities, velocity gradients and
    morphology suggest that the ionized gas is physically related to this molecular cloud.

(6) The geometrical arrangement of the stellar and gaseous components in the Arched Filament region is
    complex. Over much of the eastern Arched Filaments (E1 and E2) the ionized gas appears to lie on the near edge of the molecular gas, yet in the western filaments (W2 in particular), some molecular material must lie in front of the ionized gas, consistent with HI absorption results (Lasenby et al. 1989). The narrow and curved morphology of the Arched Filaments therefore suggests that the molecular cloud has a finger-like distribution of molecular material, the edges of which are ionized. The cluster is likely to be embedded within the distribution of molecular gas such that some of the cloud surfaces along our line of sight have not been exposed to the ionizing radiation. 

\section{Acknowledgements}

We thank Anneila Sargent for assistance with our OVRO millimeter array
observations. We also thank Debra Shepherd (and the rest of the OVRO staff) 
for observational support and advice at the
millimeter array, and for assistance with mosaicking and the addition of the
30-m and OVRO data. In addition, we are grateful to Gene Serabyn who kindly supplied the IRAM 30-m observations and to Francois Viallefond who provided great assistance in the imaging of the IRAM data. Research at the Owens Valley Radio Observatory is supported by the National
Science Foundation through NSF grant number AST 9981546.

\clearpage

\begin{figure}
\plotone{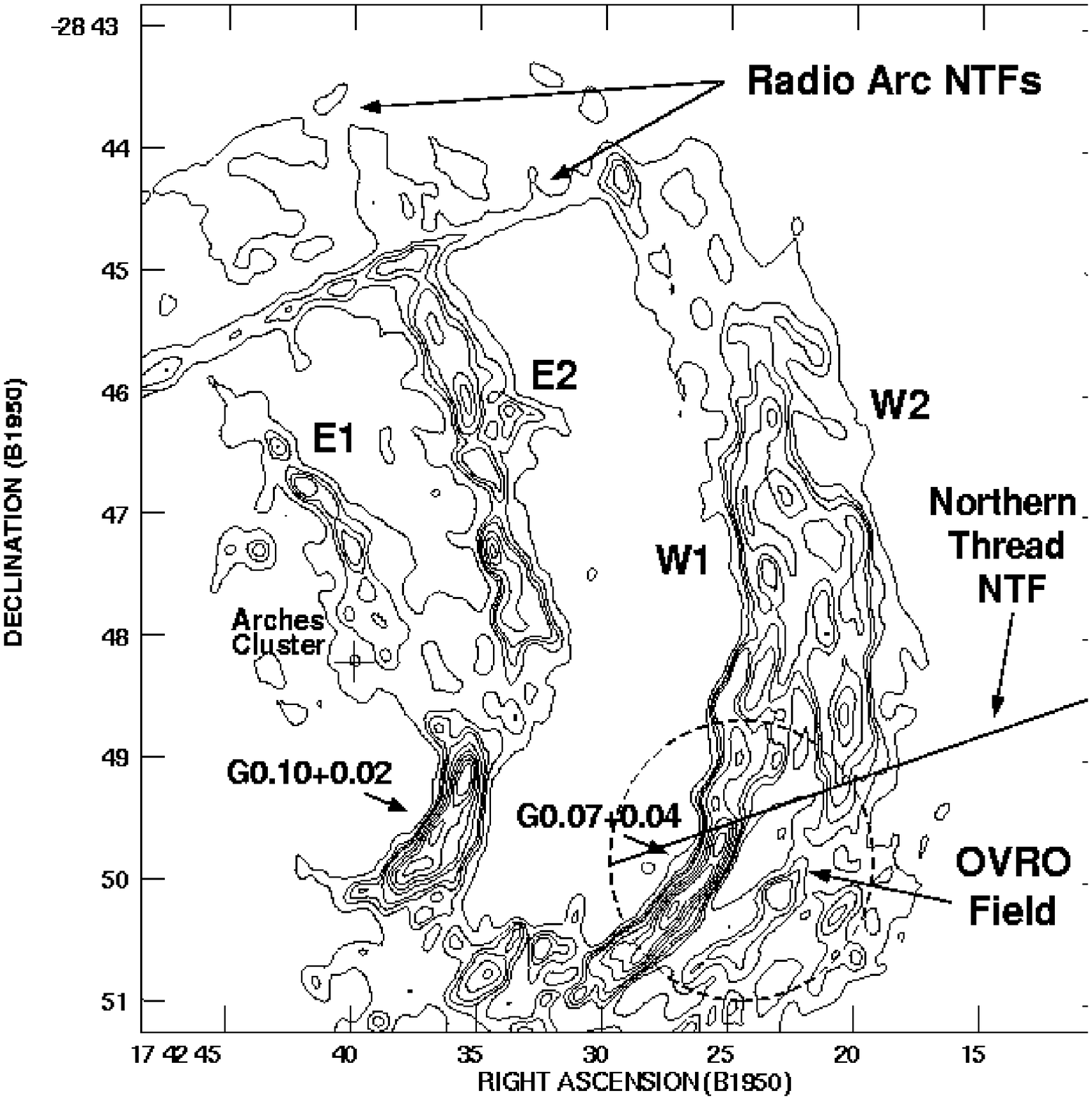}
\caption{8.3 GHz continuum image of the Arched Filaments H II complex (with resolution of 7\farcs8 $\times$ 6\farcs7, PA=$-$1\fdg0) from LGM01 showing the different components of the Arched Filaments (E1, E2,W1, W2, G0.10+0.02 \& G0.07+0.04), the Radio Arc and Northern Thread NTFs, and
the field of view of the OVRO millimeter array observations. In addition, the
position of the Arches stellar cluster is indicated by a cross.}
\end{figure}

\begin{figure}
\plotone{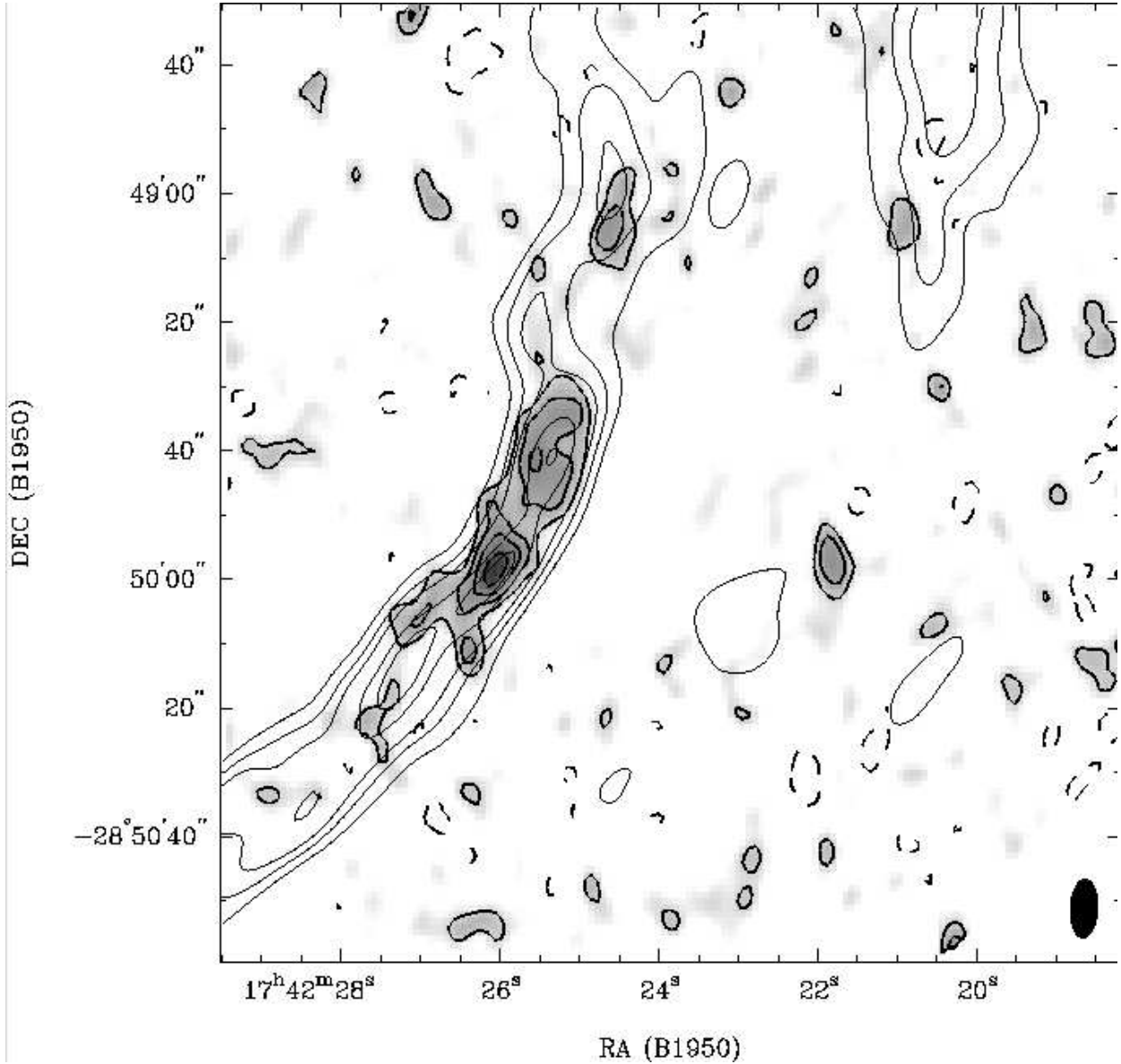}
\caption{3.4 mm continuum emission from mosaicked OVRO millimeter array
observations of the G0.07+0.04 H II region. Greyscale and darker contours show
3.4 mm continuum emission at levels of 13, 19, 26, 32, and 35 mJy \beam, with a
resolution of 9\farcs2 \x~4\farcs6, PA=$-$3\fdg2. This image has been corrected
for primary beam attenuation. The lighter contours represent VLA 8.3 GHz (3.6
cm) continuum emission in the W1 and W2 filaments with resolution of 7\farcs8 $\times$ 6\farcs7, at levels of 24, 36, 48, 72, \& 84 mJy \beam~from LGM01.}
\end{figure}

\begin{figure}
\plotone{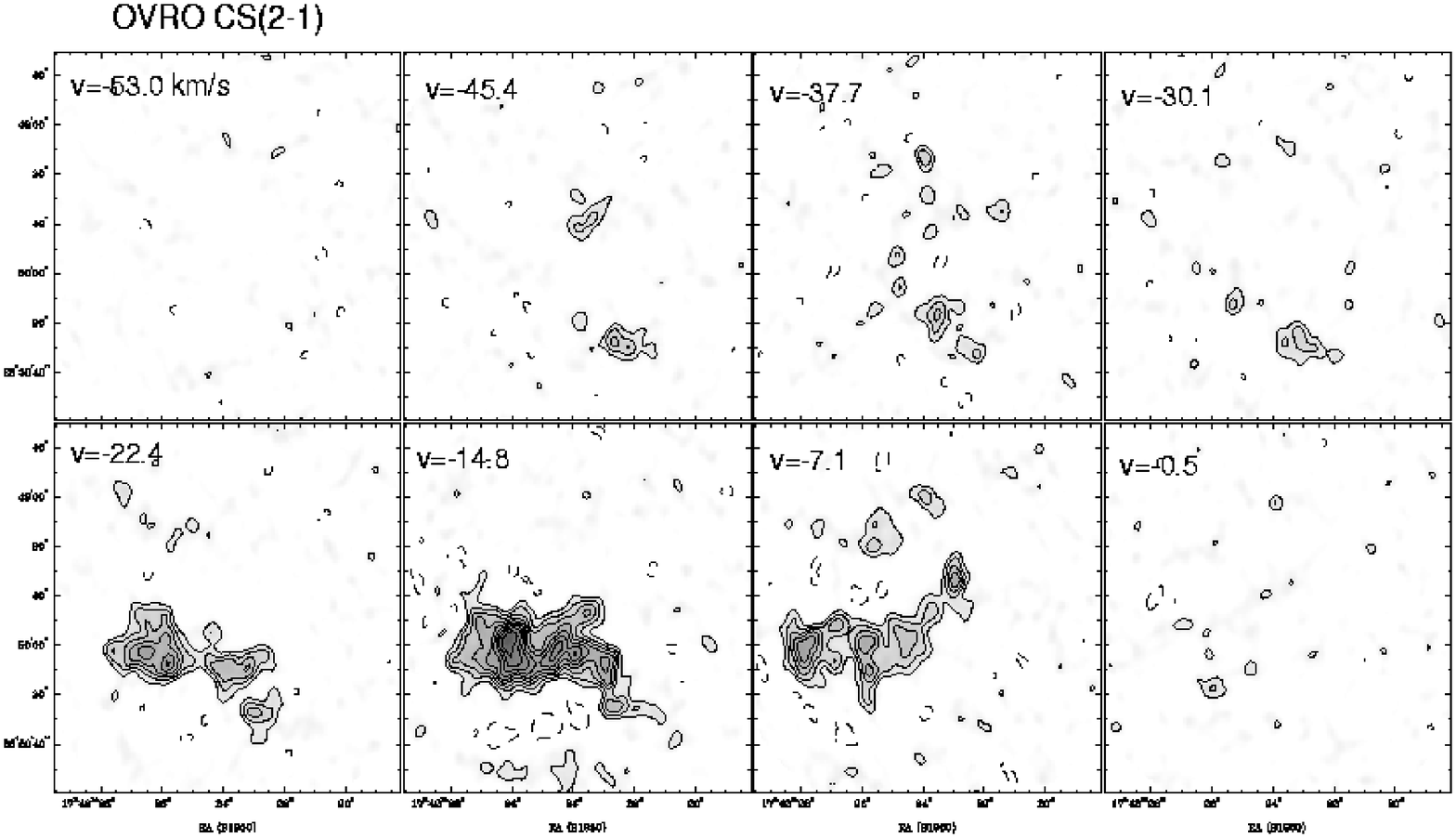}
\caption{CS(2-1) emission from OVRO millimeter array observations. These channel
images show the line emission in the range of $-$53 to $-$0.5 \kms, with each
image representing a sum over 5 channels, or \ab7.5 \kms~in
velocity. The resolution of the images is 8\farcs1 \x~4\farcs9, PA=$-$10\fdg8. 
The peak in the CS line (1.1 Jy \beam) occurs at \ad=17 42 26.0, $-$28 50 00,
V$_{LSR}$=$-$14.8 \kms.  The contours and greyscale span the range 0 to 55 mJy
\beam.}
\end{figure}

\begin{figure}
\plotone{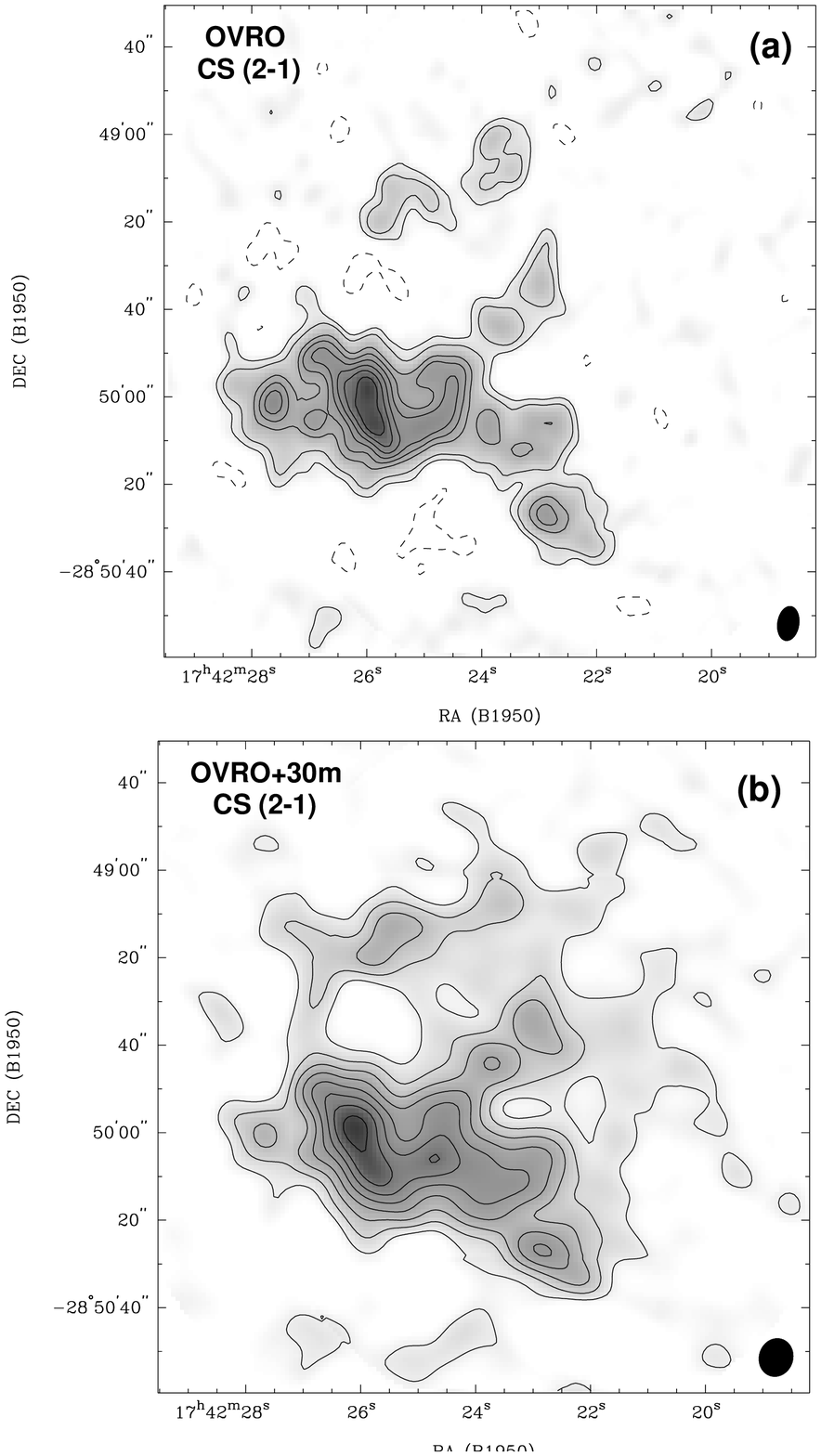}
\caption{CS(2-1) emission integrated over the range of 0 to $-$55 \kms~from (a)
OVRO only observations with a resolution of 8\farcs1 \x~4\farcs9, PA=$-$10\fdg8
and (b) OVRO+30-m data which has a resolution of 8\farcs~\x~7\farcs8,
PA=$-$17\fdg8.}
\end{figure}

\begin{figure}
\plotone{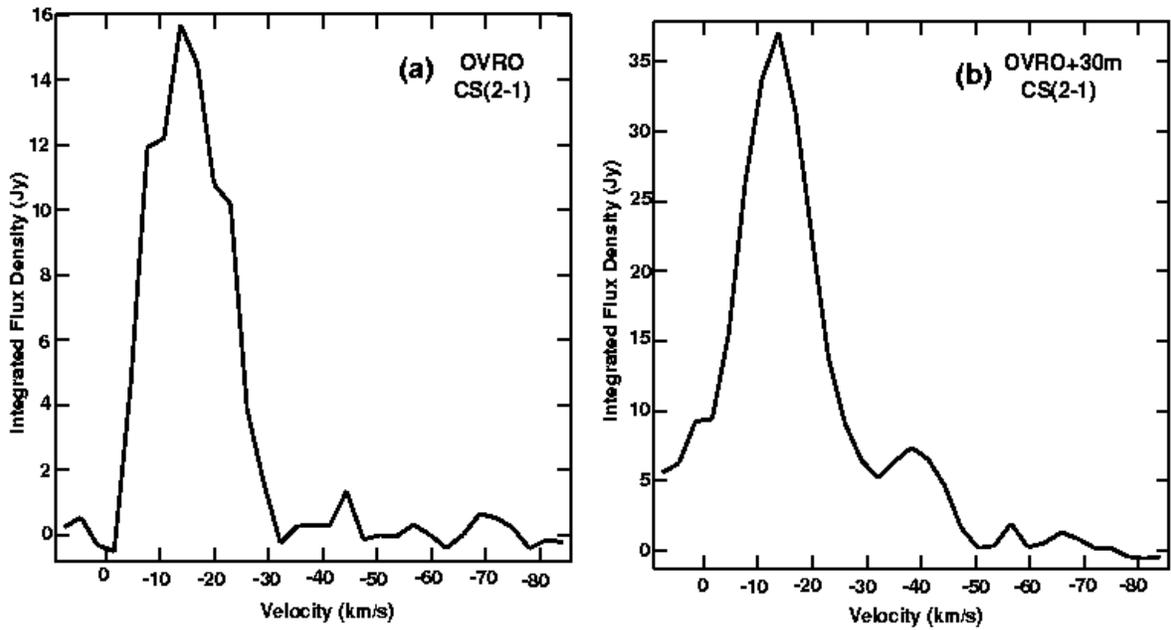}
\caption{CS(2-1) spectra integrated over a 1\arcmin~\x~1\arcmin~region
surrounding the peak emission from (a)the OVRO only data and (b) the OVRO+30-m
data. The missing flux density in the interferometer data is apparent in comparison with
the OVRO+30-m spectrum.}
\end{figure}

\begin{figure}
\plotone{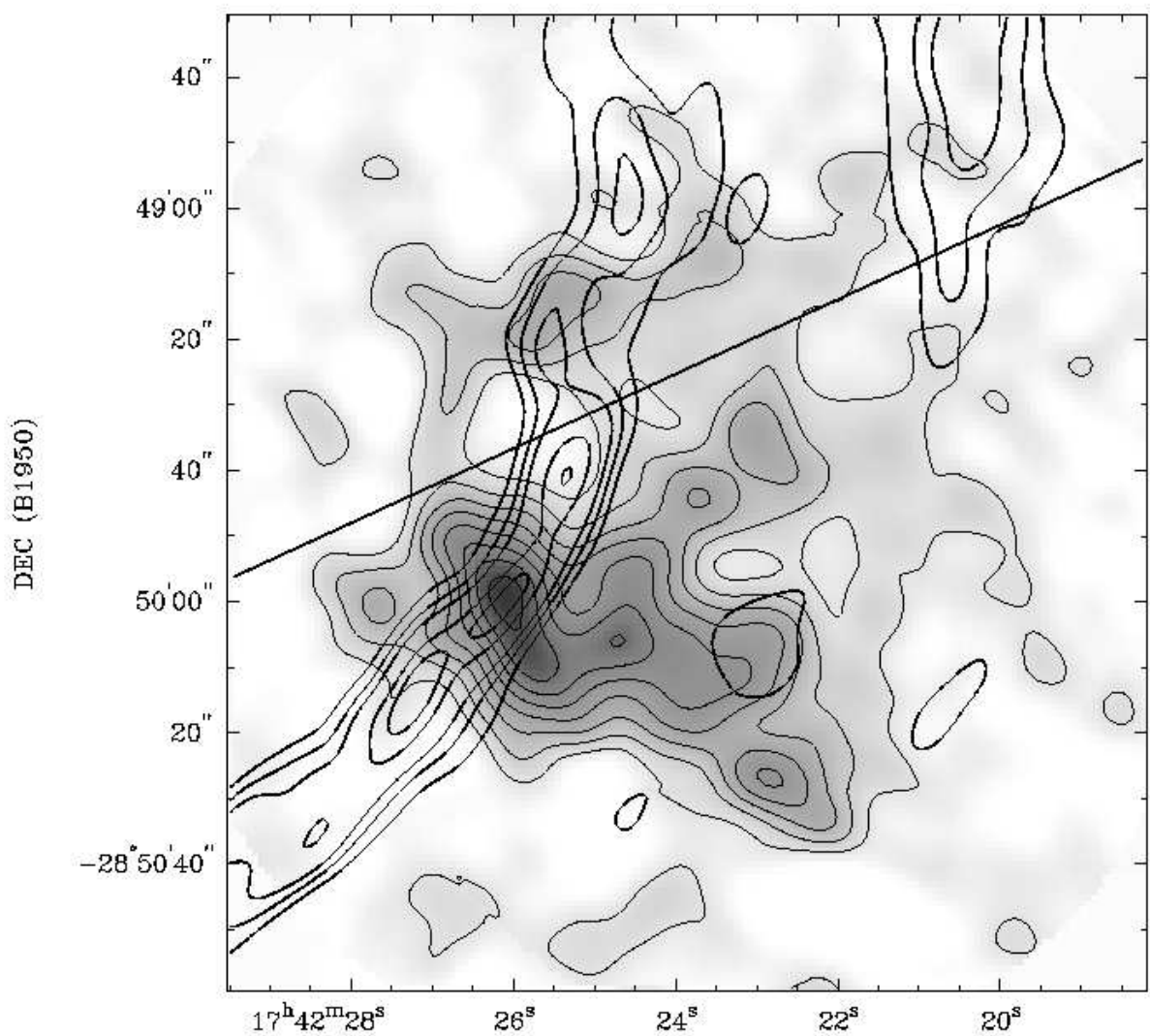}
\caption{CS(2-1) emission from the OVRO+30-m data, integrated over the range of
0 to $-$50 \kms~shown with contours and greyscale as in Figure 4b. The darker
contour overlay represents the VLA 8.3 GHz emission arising from the W1 and W2
Arched Filaments at levels of 24, 36, 48, 72, \& 84 mJy \beam, with a resolution
of \ab8\arcsec~from LGM01. The line represents the position of the Northern Thread NTF.}
\end{figure}

\begin{figure}
\plotone{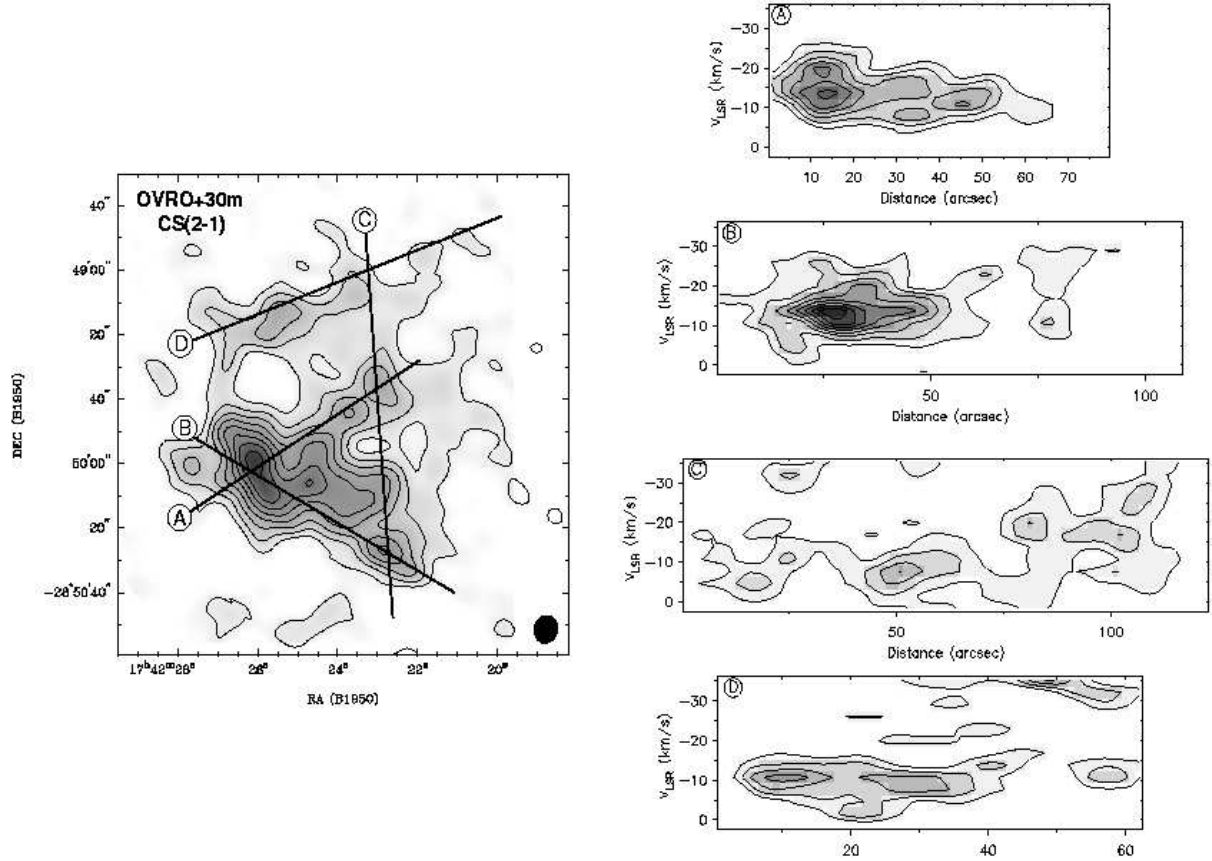}
\caption{(left panel) Integrated CS emission from OVRO+30-m data as in Figure 4b, with
lines representing the four position angles along which the position-velocity
diagrams were made. The circled letters (ABCD) represent the starting positions. (right panels)
Position-velocity diagrams of the CS emission along cuts ABCD: in panels A and B, the contour
levels represent the range 0.50 to 2.5 Jy \beam, at intervals of 0.25 Jy \beam,
and in panels C and D, the contours correspond to 0.25 to 2.5 Jy \beam~with the same
interval.}
\end{figure}

\begin{figure}
\plotone{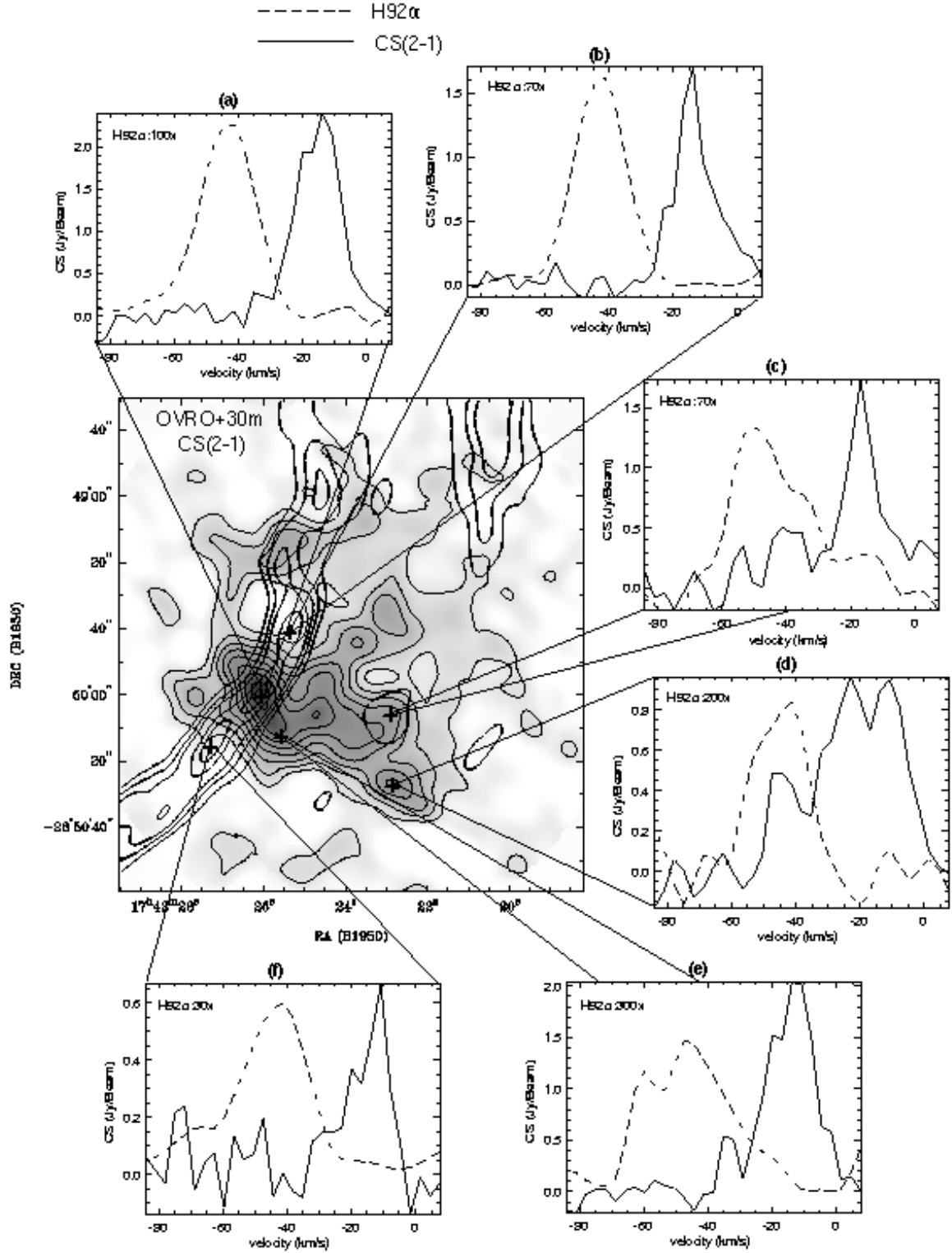}
\caption{A comparison of ionized and
molecular gas in G0.07+0.04. The greyscale shows integrated CS(2-1) emission (OVRO+30-m) and contours of 8.3 GHz continuum. Solid lines represent the line profiles of CS(2-1) emission, and the dotted lines the H92$\alpha$ line from the same positions (LGM01).}
\end{figure}

\begin{figure}
\plotone{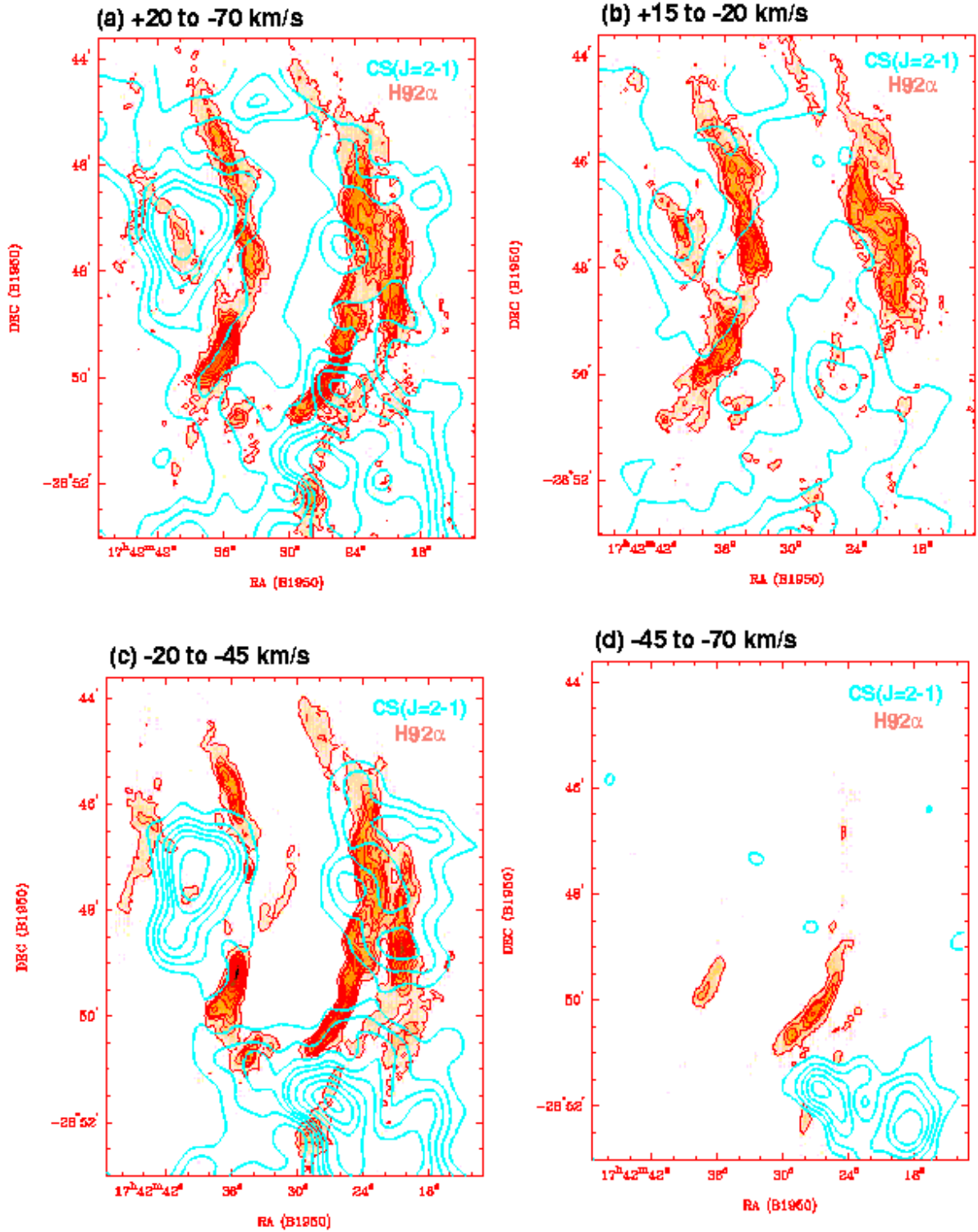}
\caption{A comparison of the integrated emission from the IRAM 30-m CS(2-1) data
of SG87 shown in turquoise contours and the H92$\alpha$
line emission LGM01 shown in colorscale and contours. The region
covers the Arched Filaments and the four panels show the emission integrated
over different velocity ranges which are labelled above each figure.}
\end{figure}

\begin{figure}
\plotone{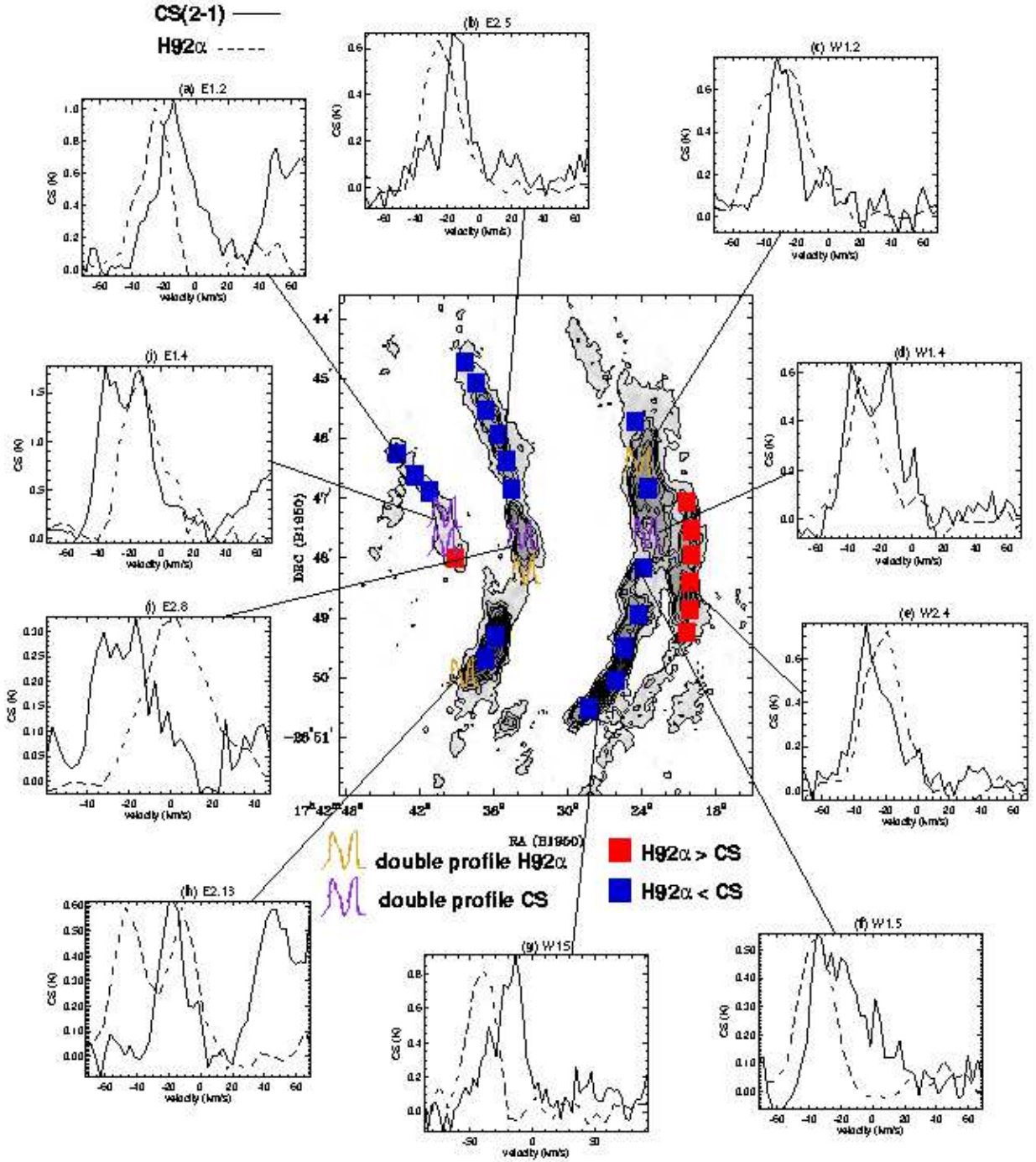}
\caption{A comparison of the velocities and lineshapes of the ionized and
molecular gas across the entire Arched Filaments/$-$30 \kms~cloud region. In each of the panels (labelled a-j) the
solid line profiles represent the CS(2-1) emission from the data of SG87, and the dotted line profiles show the H92$\alpha$ line emission sampled at the same positions from LGM01 which are scaled by an arbitrary
factor to fit on the scale of the CS profiles. The colors represent the relative
differences between the velocities of the ionized and molecular components (i.e. whether the ionized gas is redshifted or blueshifted with respect to the molecular gas). The
central region represents the integrated H92$\alpha$ emission from LGM01.}
\end{figure}

\begin{figure}
\plotone{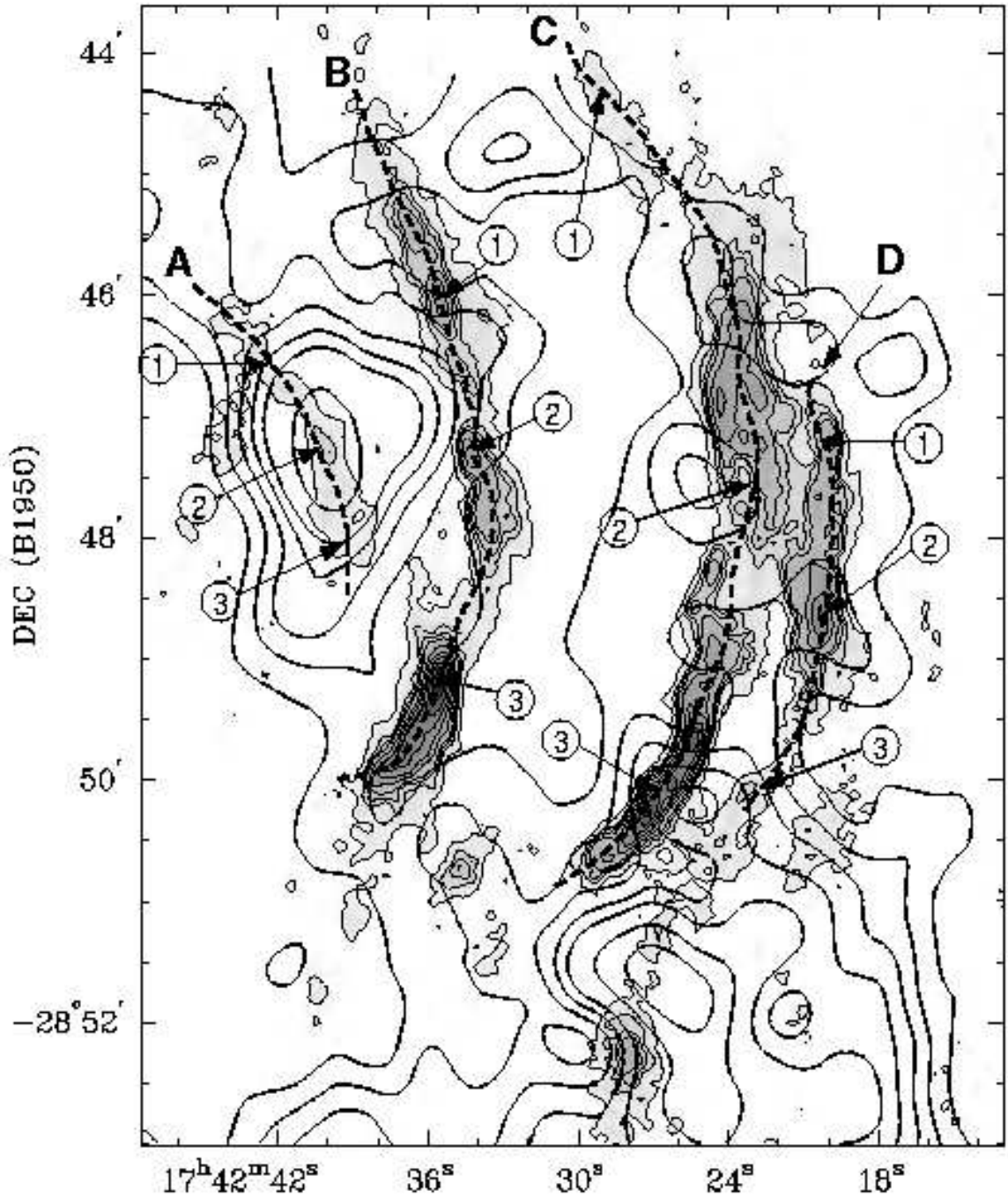}
\caption{Finding chart for the positions at which position-velocity diagrams
have been constructed for the CS(2-1) emission from the observations of Serabyn
\& \gusten~(1987) and the H92$\alpha$ lines of LGM01. The dark contours here represent the distribution of integrated CS(2-1) emission and the greyscale and lighter contours represent the distribution of integrated H92$\alpha$ emission (as shown in Figure 9a). The labels ABCD correspond to the four filaments E1, E2, W1 and W2 and the panels of Figure 12 (a-d). The dotted lines represent the positions where the velocity was sampled along each filament. The numbers (1-3) along each filament serve as reference markers for the diagrams shown in Figure 12.}
\end{figure}

\begin{figure}
\plotone{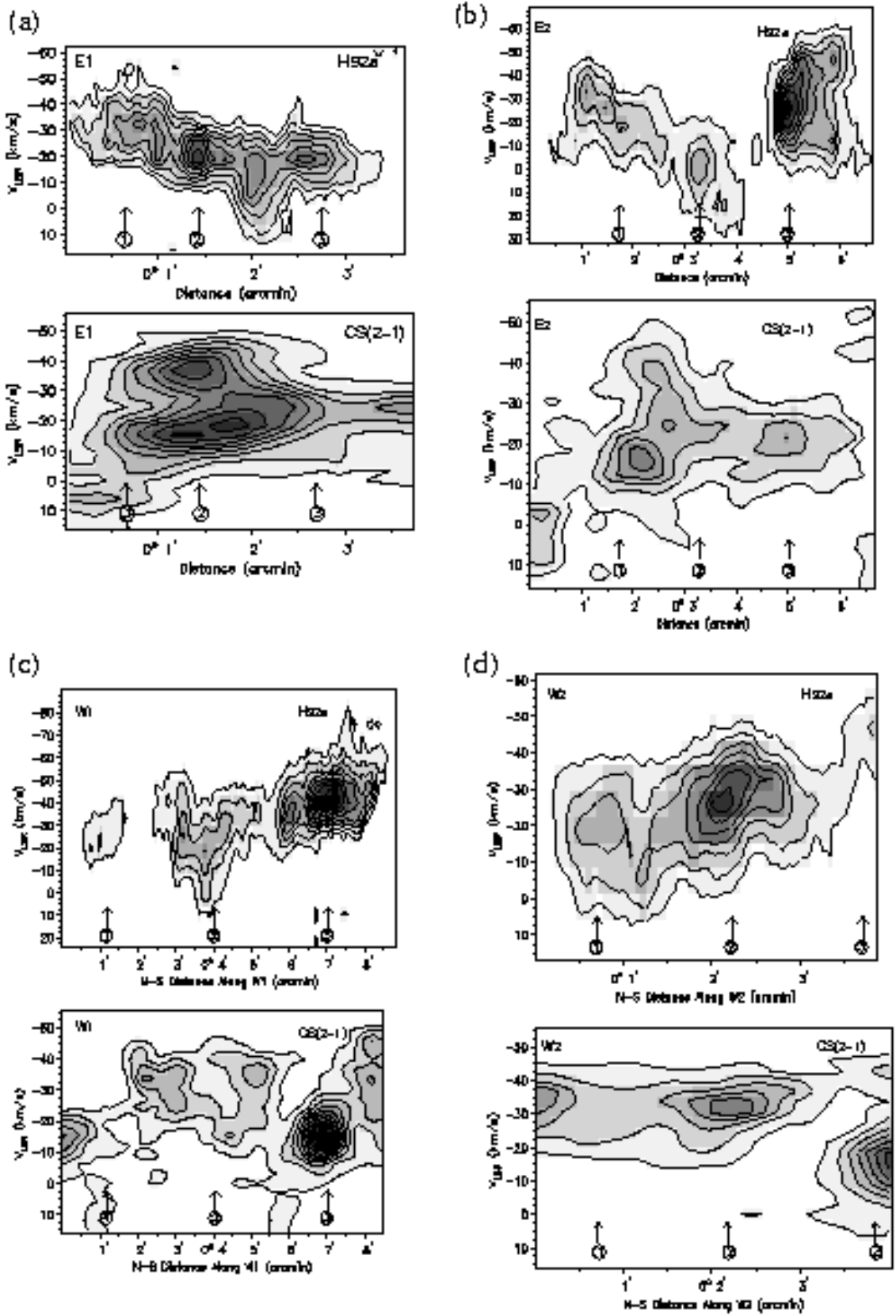}
\caption{Position-velocity diagrams for the four Arched Filaments as in Fig. 11.}
\end{figure}

\begin{figure}
\plotone{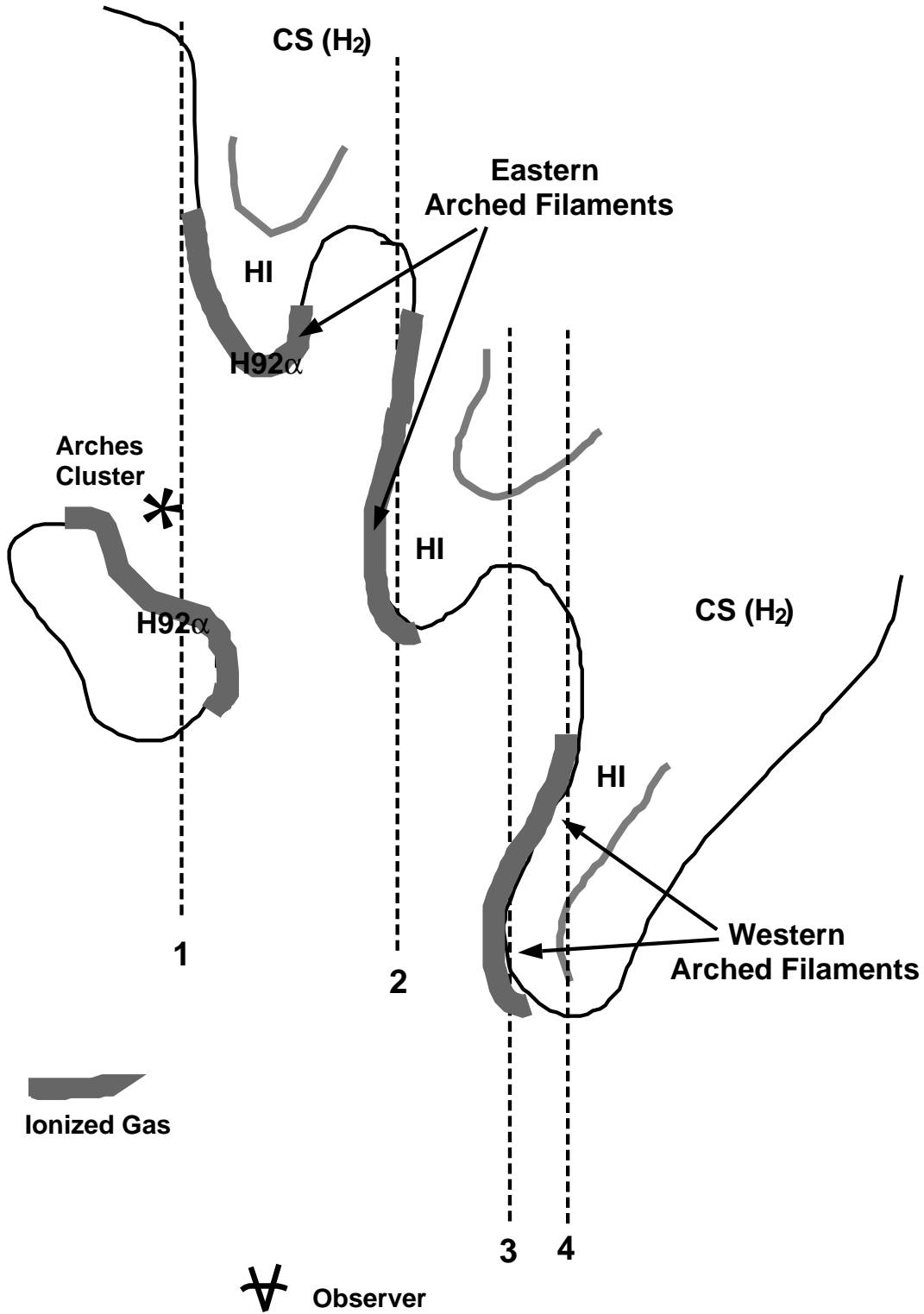}
\caption{A schematic diagram showing a view of the ionized, molecular, atomic, and stellar components in the Arched Filament complex from a position ``above'' the molecular cloud, looking down its long axis.}
\end{figure}

\end{document}